# Dark Matter, Quasars, and Superstructures in the Universe (with $\delta$-particle search and spherical universe)

Xiaodong Huang
*Department of Mathematics, University of California, Los Angeles
Los Angeles, CA 90095, U.S.A.*

Wuliang Huang
*Institute of High Energy Physics, Chinese Academy of Sciences
P.O. Box 918(3), Beijing 100049, China*

**Abstract:** From the observed results of the space distribution of quasars we deduced that neutrino mass, $m_\nu$, is about $10^{-1} eV$. The fourth stable elementary particle $\delta$ with mass $m_\delta \sim 10^0 eV$ can help explain the energy resource mechanism in quasars, cosmic ultra-high energy particles, as well as the flatness of spiral galaxy rotation curves. The blue bump and IR bump in the quasar irradiation spectra, as well as the peaks of EBL (Extragalactic Background Light) around $10^0 eV$ and $10^{-1} eV$, are related to the annihilation of $\delta\bar{\delta}$ and $\nu\bar{\nu}$ respectively. This enlightens us to explore the reason for missing solar neutrinos and the unlimited energy resource in a new manner. For $\delta$-particle search it is related to Dual SM or Two-fold SM; the relationship between space electron spectrum ($>10^0 TeV$) and cosmic ray spectrum (knee and ankle) at high energy region; and the characteristics of spherical universe. Appendix is the theory part related to mass tree, inflation, BSM, finite universe etc.

**PACS numbers:** 95.35. +d, 98.65.Dx, 98.54.-h, 26.65.+t, 89.20.-a

**Keywords:** dark matter, quasar, superstructure, new energy resource, cosmic ultra-high energy particles, mass tree, inflation, beyond SM, finite universe (spherical universe)

______________________________________________________________
Email address: huangwl39@yahoo.com, huangwl@ihep.ac.cn, xhuang@ucla.edu

## Superstructure and Neutrino Mass

In Refs [1],[2] and [3], we mentioned that there exist dark matter particles $\nu$ ($\bar{\nu}$) with mass $m_\nu(m_{\bar{\nu}}) \sim 10^{-1} eV$, and may also exist dark matter particles $\delta$ ($\bar{\delta}$) with mass $m_\delta(m_{\bar{\delta}}) \sim m_\nu(m_{\bar{\nu}})$. The dark matter particles involve the superstructure with mass scale $M_F \sim 10^{19}$ solar mass and with length scale (at present time) $r_F \sim 10^3$ Mpc in the universe. Evidence pertaining to this topic is appeared: In Ref [4], from the quasar number count as a function of redshift, $N(z)$, there are several peaks between $z = 0$ and $z = 5$. There are strong peaks in $N(z)$ at $z \approx 0.24$, 1.2, and 1.8.[4] Using the results of the Fourier spectral analysis in that paper, we roughly adopted the redshift values of the peaks as follows:

| i | 1 | 2 | 3 | 4 | 5 | 6 |
|---|---|---|---|---|---|---|
| $z_i$ | 0.24 | 0.5 | 0.8 | 1.2 | 1.8 | 3.0 |
| $d_i$ (Gpc) | 0.86 | 1.55 | 2.15 | 2.75 | 3.40 | 4.22 |
| $\Delta d_i$ (Gpc) | 0.86 | 0.69 | 0.60 | 0.60 | 0.65 | 0.82 |

Where $d_i$ is the distance, $\Delta d_i = d_i - d_{i-1}$ and $d_0 = 0$. The average value of $\Delta d_i$ is 0.7 Gpc ($\tilde{<} 10^3 Mpc$).** This is the length scale of superstructure in the universe. Therefore, it inverses to deduce that $m_\nu(m_{\bar{\nu}}) \sim 10^{-1} eV$ as in Refs [1] and [2], but it is not certain whether $m_\delta > m_\nu$ or $m_\delta < m_\nu$. In this paper, we are still using the order of magnitude estimation as in [1] and [2], and denote dark matter particles ($\nu$ and $\delta$) as "$d$" and baryon particles as "$B$".

## S galaxies and quasars

In Ref [6], we discussed galaxies with two constituents: dark matter and baryon matter ($d + B$, $m_d \ll m_B$). In that paper, we introduced a parameter "n" which can be used to distinguish spiral (S) galaxy and elliptical (E) galaxy: n > 0 for S galaxies; n < 0 for E galaxies. From calculation, it is obvious that a galaxy contains more dark matter if n > 0. Therefore, S galaxies contain more dark matter than E galaxies. Comparing number count functions $N(z)$ of S and E galaxies[7] with quasars,[4] one can see that the $N(z)$ function of S galaxies is more similar to that of quasars. Thus, one of the quasar's characteristics is that it contains abundant dark matter (and anti-dark matter).

On the other hand, the fact that $N(z)$ of both S and E galaxies have several peaks[7] means that both types of galaxy do not have a homogeneous space distribution in a superstructure. S and E galaxies may occupy outer region or central region respectively. In a superstructure, the light constituent d is enriched (with larger abundance rate [8]) in outer regions and the heavy constituent B is enriched in the central region. This means that S galaxies and quasars are mainly distributed in the outer region of a superstructure.

---
** Using the $N(z)$ data for quasi-stellar objects in Ref [5], we have a result of average $\Delta d_i$ ~0.6 Gpc.

## $\delta$ particles and quasars

Since quasars contain abundant dark matter, the energy source of quasar can be related to dark matter. Thus, another energy resource mechanism is added to the standard black hole-accretion disk model of quasars.[9],[10] Since there are super-strong gravitational field and super-strong electromagnetic field in quasars, dark matter and anti-dark matter are annihilated in super-strong gravitational / electromagnetic fields: $\delta + \bar{\delta} \rightarrow$ graviton + graviton / $\nu + \bar{\nu} \rightarrow \gamma + \gamma$. The interaction between gravitons and the surrounding baryon matter produces blue light[11], while the annihilation of $\nu\bar{\nu}$ produces IR photons. These two types of radiations are respectively related to the "blue bump" and "IR bump"[12],[13],[14] of quasars.

The luminosity $L_q$ of a quasar is about $10^{47}$ erg/sec. If the length scale of quasars is $r_q$ (<1pc), then the intensity flowing out ($erg/\sec/cm^2$) of a quasar's surface is $j_q \sim \frac{L_q}{r_q^2}$.

At the surface, the number density of d-particle ($n_d$) can be estimated as $(\frac{m_d c}{\hbar})^3$. During the annihilation process, the intensity entering the surface is $j_d \sim n_d (m_d c^2) \cdot c$. It is obvious that $j_d \sim \xi \cdot j_q$ and $\xi < 1$. So we have $m_d \sim (\frac{\xi \cdot L_q \hbar^3}{r_q^2 c^6})^{1/4} \sim 10^0 eV$, i.e. $m_\delta \sim 10^0 eV$.

Therefore, $m_\delta > m_\nu$.

We discussed the mass of dark matter particles in Ref [6]. The massive neutrinos were used to explain the flatness of spiral galaxy rotation curves. From the calculation of 21 samples of S galaxies, the calculated values of neutrino mass are in the range of 5eV – 33eV.[6] According to the analysis in this paper, the dark matter particles that cause the "flatness" are not the neutrinos, but rather the $\delta$ particles.

## Cosmic Blue Light

The annihilation processes of $\delta\bar{\delta}/\nu\bar{\nu}$ may not only occur in quasars, but also in any places with super-strong gravitational field / electro-magnetic field in the universe, such as galactic nucleus, neutron star, central region of star, black holes, etc. Gravitons, the product of $\delta\bar{\delta}$ annihilation, can travel a long way in space and interact with baryon matter anywhere to emit cosmic blue light. This means the $10^{14-15}$ Hz blue light and $10^{13-14}$ Hz IR radiation can occur at different z across the entire sky and make imprints on the EBL (Extragalactic Background Light). Thus, the peaks of EBL around $10^0 eV$ and $10^{-1} eV$ (a small peak) are related to the annihilation of $\delta\bar{\delta}$ and $\nu\bar{\nu}$ respectively.[15],[16]

## Solar Neutrinos and Energy Resource

As discussed above, the annihilation processes of $\delta\bar{\delta}$ and $\nu\bar{\nu}$, especially the $\delta\bar{\delta}$ annihilation process ($m_\delta >> m_\nu$), might provide a part of solar energy. This might be a reason for "missing" solar neutrinos. Since the 'blue bump" and "IR bump" in the solar spectrum are not so clear, it is uncertain that the "missing solar neutrino" can be the result of this newly postulated additional energy source. So, we suggest another process: a portion of the solar neutrinos $\nu_e$ is annihilated with dark matter anti-neutrinos $\bar{\nu}_e$ at the central region of the sun and "missed".

If we plan to examine the annihilation processes of $\delta\bar{\delta}$ and $\nu\bar{\nu}$ in a laboratory, the first task is to create equipment that can produce super-strong gravitational field, which may be similar to a mini NIF (National Ignition Facility). Since dark matter exists everywhere, it can be automatically accreted from the outside of the equipment, thus providing an unlimited source of fuel. The products (blue light…) of this equipment can be directly transferred to electric power as solar energy equipment. So, this experiment equipment can become a new generation of artificial energy resource. Suppose the radius of reaction region is $r_c \sim 10^{-1}$cm, and $m_\delta \sim 10^0$eV, using the equation: $(m_\delta)^4 \sim \frac{\xi \cdot L_w \hbar^3}{r_c^2 c^6}$, the maximum output power (luminosity) $L_w \sim 10^4$ kw. This is a type of middle scale energy station, which can be used for individual living or spaceflight (if $m_\delta \sim 10^1 eV$, then $L_w \sim 10^8$ kw).

When accretion disks are consumed in quasars, it is possible that some point sources with "mono-color" gravitational wave / infrared emission can be observed across the sky. This will help us explore dark matter particles' mass and the energy resource mechanism in quasars.

## Search for $\delta$ particle

(1) According to Dual SM/two-fold SM (see Appendix), $\delta$ particle is like neutrino but with baryon number. We may use high energy protons (>0.5 TeV) to collide carbon/beryllium target to produce $\delta$ particles; or refer equipment that was used to search for cosmic neutrinos.

(2) For the ultra-high energy primary cosmic ray spectrum (UEPCRS), the "knee" may related to the interaction between proton and $C\nu B$ ( $p + \nu/\bar{\nu} \to n + e/\bar{e}$ ) and the "ankle" may related to the interaction between proton and $C\delta B$ ( $p + \delta/\bar{\delta} \to n + p/\bar{p}$ ). If space electron spectrum is correlated with UEPCRS, it will appear two abnormalities by $10^0 TeV$ and $10^2 TeV$. The latter is related to $\delta$ particle. Besides, for cosmic ultra-high energy particles (such as $\nu_\mu$ [17]), the key is the production of ultra-high energy neutrons, which may create in the interaction process of $\delta$ and $p$.

(3) We may live in a spherical universe as a u-particle and the earth locates near the center (see Appendix). Then, the CMB anisotropy spectrum may reflect the mass spectrum of $\nu$ particles and $\delta$ particles. And the CMB anisotropy spectrum itself is "anisotropic".

(4) Basis on mass tree (see Appendix) it is prefer to look for $\delta$-particles in galaxy-clusters.

# Appendix
The detail is in Ref [3] and Ref [2].

**Large number and mass scales sequence (mass tree)**

In our universe, the fundamental physical constants are the speed of light $c$, the gravitation constant $G$, and the Planck constant $\hbar$; the fundamental block of the mass is the most stable baryon proton with mass $m_p$ and radius $r_p$ ($r_p \sim \frac{\hbar}{m_p c}$).

From $c, G, \hbar$, we have

$$\text{Planck mass } m_{pl} \sim \sqrt{\frac{\hbar c}{G}} \sim 10^{19} GeV \qquad (1)$$

$$\text{Planck length } r_{pl} \sim \sqrt{\frac{\hbar G}{c^3}} \sim \frac{G m_{pl}}{c^2} \qquad (2)$$

then
$$\text{Large Number } A \sim \frac{m_{pl}}{m_p} \sim \frac{r_p}{r_{pl}} \sim 10^{19} \qquad (3)$$

The mass scales sequence of the universe was suggested twenty years ago, now it can have the diagram (mass tree) as follows:

$$m_0 = M_0$$
$$m_1 \qquad M_1$$
$$m_2 \qquad M_2$$
$$m_3 \qquad M_3$$
$$m_4 \qquad M_4$$
$$m_5 \qquad M_5$$
$$m_6 \qquad M_6$$
$$m_7 \qquad M_7$$
$$m_8 \qquad M_8$$

On the left side of the diagram are the mass scales of stable particles in micro-cosmos:

$$m_n = A^{-\frac{n}{2}} m_{pl}, \text{n} = 0, 1, 2 \ldots \qquad (4)$$

(with length scale $r_n$)

On the right side are the corresponding mass scales of celestial bodies in macro-cosmos:

$$M_n = A^n m_{pl}, \text{n} = 0, 1, 2, 3\ldots \qquad (5)$$

(with length scale $R_n$ and density $\rho_n$)

For n=0, $m_0 = M_0 = m_{pl}$

For n=1, $m_1 = A^{-0.5} m_{pl} \sim 10^{18} eV$ (super heavy particle)

$\qquad M_1 = A \cdot m_{pl}$ (lightest black hole - LBH)

For n=2, $m_2 = A^{-1} m_{pl} = m_p$ (proton)

$\qquad M_2 = A^2 m_{pl} = A^3 m_p = M_{star}$ (star)

For n=3, $m_3 = A^{-1.5} m_{pl} = A^{-0.5} m_p = m_\nu \sim 10^{-1} eV$ (neutrino)

$\qquad M_3 = A^3 m_{pl} = A \cdot M_{star}$ (superstructure)

For n=4, $m_4 = A^{-2} m_{pl}$

$\qquad M_4 = A^4 m_{pl} = A^3 M_1$ (end of inflation, critical energy $E_{cr}$)

For n=5, $m_5 = A^{-2.5} m_{pl}$

$\qquad M_5 = A^5 m_{pl} = A^3 \cdot M_{star} = M_u$ (u-particle, the beginning of inflation)

.
.
.

For n=8, $m_8 = A^{-4} m_{pl} = A^{-3} m_p = m_A$ (A-particle, the most elementary particle)

$\qquad M_8 = A^8 m_{pl} = A^3 \cdot M_u$ (original universe)

From above diagram, there is a main mass sequence in the universe:

$m_8 \rightarrow m_2 \rightarrow M_2 \rightarrow M_5 \rightarrow M_8$

i.e. $m_A \times A^3 = m_p$; $m_p \times A^3 = M_{star}$; $M_{star} \times A^3 = M_u$; $M_u \times A^3 = M_8$.

It is obvious that the large number "$A$" plays an important role in both micro-cosmos and macro-cosmos as a fundamental physical constant. In cosmology, the fundamental physical constants are $G, c, \hbar, m_p$ or $G, c, \hbar, A$.

For n=0,1,2 that represent the fundamental blocks in the universe:
n=0, $M_0 = m_{pl}$, $R_0 = r_{pl} \sim \lambda_0$; $\lambda_0$ reflects the fundamental block of space-time;
n=1, $M_1 = A \cdot m_{pl} = M_{LBH}$, it is the fundamental block of the early universe;
n=2, $M_2 = A^2 m_{pl} = M_{star}$, it is the fundamental block of the visible universe.
For n=3,4,…8 that represent the evolution of the universe before H-decoupling;
From $R_5$ to $R_8$ all have a minimum radius $r_{min}$ ($R_5 = R_6 = R_7 = R_8 = r_{min}$) without singularity at the beginning of time. $m_4$, $m_5$, $m_6$, $m_7$, $m_8$ are background particles.

Our universe is a u-particle as a spherical universe.
We can insert mass scale between $M_2$ and $M_3$: $M' \sim M_{starcluster}$, $M'' \sim M_{galaxycluster}$;
The corresponding masses of particles are $m' = m_e$ (electron) and $m'' = m_\delta$ ($\delta$ particle).

According to observation cosmology the evolution procession (equations) of spherical universe is like SM obeyed the cosmology principle except in the early era ($> E_{cr}$).

**Critical density $\rho_{cr}$ and lightest black hole (LBH)**

The density of the universe is $\rho_{univ}$. If we suppose that $\dfrac{\rho_{univ}}{c^2} = const.$ for $\rho_{univ} > \rho_{cr}$, our universe will naturally have an inflation stage ($\dot{R}^2 \propto R^2$).

[$\dfrac{\rho}{c^2} = const.$ for $\rho > \rho_{cr}$] means that there is a minimum radius $r_{min}$ for all $\rho > \rho_{cr}$ black holes. We have

$$r_{min} \sim c/\sqrt{G \cdot \rho_{cr}} \tag{6}$$

and the mass of LBH $\quad M_{LBH} \sim r_{min} \cdot c^2 / G \tag{7}$

Suppose $\quad r_{min} = r_p \tag{8}$

Then $\quad M_{LBH} \sim r_p \cdot c^2 / G = A \cdot m_{pl} = M_1$, $R_1 = r_p \tag{9}$

From eqs (6), (7), (8) $\quad \rho_{cr} \sim A^2 \cdot \rho_p$ and $\rho_p \sim \dfrac{m_p}{r_p^3}$. $\tag{10}$

**Critical Energy, Inflation, and TOE**
Imagining the inverse process of evolution of the universe, when $\rho_{univ}$ arrive at $\rho_{cr}$ (n=4) from present CMB density ($\rho_{CMB}$): $M_4 \sim A^4 m_{pl} \sim A^3 M_1$, $R_4 = R_{star} \sim A \cdot R_1$ and $\rho_4 \sim \dfrac{M_4}{R_4^3} = \dfrac{M_1}{R_1^3} \sim \rho_1 = \rho_{cr}$. It means that, at this time, our universe includes $A^3$ lightest black holes. Since then, the LBHs were merged and collapsed to a "u particle", which has mass scale $M_u = M_5 \sim A^5 \cdot m_{pl}$ and length scale $R_5 \sim r_p$. It is obvious that the positive process of evolution from $M_5$ to $M_4$ be exactly the inflation process of the universe. So, the inflation is a step by step fission process of black holes (more and more LBH appear) and the CMB may have a fine grain structure.

The energy scale at $R_4$ is the critical energy $E_{cr}$, from $\rho_{cr}$ we have

$E_{cr} \sim (m_p^2 c^9 \hbar G^{-1})^{1/4} \sim 10^{18} eV$ , which is like the cutoff of renormalization. At $E_{cr}$, when LBH are produced, the boundaries of elementary particles for SM or BSM in our spherical universe are disappeared. Without skin, the hairs adhere where? So, the interacted fields are also "disappeared" (unified). We will confront five back ground particles/field: $m_4, m_5, m_6, m_7, m_8$ in big universe (see Discussion-5). This is the TOE in a new manner.

We can select different $r_{min}$ and different n-sequence. For example $r_{min} = r_{pl}$, n= $0, \dfrac{1}{2}, 1, \dfrac{3}{2}, 2 \ldots 8$, but the frame of the results is not change.

**BSM, Dark matter particles with low mass**
The diagram of mass scales sequence is like a "mass tree". The diagram of different SM of particle physics is like "pods" (with symmetry) on the tree.
We suggest Dual Standard Model diagram as follow:

| u | c | t | $\gamma$ | $u_l$ | $c_l$ | $t_l$ | G |
|---|---|---|---|---|---|---|---|
| d | s | b | g | $d_l$ | $s_l$ | $b_l$ | $g_l$ |
| $\nu_e$ | $\nu_\mu$ | $\nu_\tau$ | $Z^0$ | $\delta$ | $\delta'$ | $\delta''$ | $Z'$ |
| e | $\mu$ | $\tau$ | $W^\pm$ | p | p' | p'' | $W'$ |

Where $u_l$ $c_l$ $t_l$ $d_l$ $s_l$ $b_l$ are lept-quarks, $g_l$ is lept-gluon, $G$ is graviton. $Z', W'$ are the gauge bosons about a new type of interaction (relax interaction) related to $\delta$ particles. We can look for $G, p'$...etc. The speed of photon $\gamma$ is $c$. The speed of graviton $G$ is $c'$. There is a CGB like as the CMB. We also suggest Two-fold Standard Model:

| u | c | t | $\gamma$ | u' | c' | t' | G |
|---|---|---|---|---|---|---|---|
| d | s | b | g | d' | s' | b' | g' |
| $\nu_e$ | $\nu_\mu$ | $\nu_\tau$ | $Z^0$ | $\delta$ | $\delta'$ | $\delta''$ | $Z'$ |
| e | $\mu$ | $\tau$ | $W^\pm$ | p | p' | p'' | $W'$ |

In these models, new quarks may be related to new particles ($\pi'_0, K'_0, e'_+ \ldots$) with mass $10^2 GeV$-$10^1 TeV$ ; $\delta - particle$ and $\nu - particle$ are dark matter particles with low mass ($10^{-1} eV - 10^1 eV$). During the cooling process of LBH at $R_4$ (and also of the collision fire ball in laboratory), if a lot of electrons create before protons, the Dual SM is priority.

## Spherical universe

(1) GRBs Ring, multi-components universe and pancake.
Recently it was reported a giant ring-like structure with a diameter of 1.7 Gpc displayed by GRBs. This giant ring can be explained by pancake process in a (B+$\nu$) two components superstructure, $m_\nu \sim 10^{-1} eV$, and our universe is inhomogeneous. We think that the cosmology principle needs to be checked.

(2) CMB cold spot and spherical universe.
The report about CMB cold spot is for a long time. One of the possible explains is our universe is finite (spherical universe with present $R_{CMB} \sim 10^{28} cm$) and the earth is not at the center. If it is so, we can see the CMB anisotropy spectrums of cold semi-sphere is different to that of hot semi-sphere (when we make observation face/reverse the CMB cold spot, the first peak of CMB anisotropy spectrum will drop/rise while the sixth peak will rise/drop). And the peaks of CMB anisotropy spectrum may reflect the mass spectrum of $\nu$ and $\delta$ particles.

(3) The civilization layer (ball).
We live in a spherical universe. According to the size of the CMB cold spot, recently the distance we deviate from the center of the sphere (using light year) is in numerical less than the H-decoupling time (also using year).
The center region of the spherical universe is the civilization layer (ball), according to astronomical scale; it is just nearby the earth.

## Discussion

(1) The speed of light is not a constant; it has a mutation (phase transition) when $E = E_{cr}$ at the point $n = 4$ of mass tree in the early era of the evolving process of the spherical universe. At point $n = 5,6,7,8$ of mass tree that may also have different phase transitions.

(2) The speed of light represents the limit of the velocity of elementary particles. The limit speed of elementary particles belong to different category or different cosmic level ($n = 4,5,6,7,8$) is different.

(3) The microwave background (CMB) is from the atomic level, correspondingly, the gravitational wave background (CGB) is from nuclear level. In the collision of electrons and positrons to find photons, then in the collision of protons and anti-protons (or deuterons and anti-deuterons) gravitons can be found.

(4) The main components of the dark matter are neutrino and $\delta$ particle, which is similar to the inert neutrino but with baryon number and related to new interaction. They are expressed in a dual/two-fold SM, and make up the superstructures in our universe. $M_3$- superstructures are latent and in taking shape structures, but they speed up the formation of the sub level celestial bodies.

(5) It is possible that spherical universe itself does not have dark energy, which is the superstructure effect. However, dark energy may associate with the background particles/field of the big universe in which our spherical universe is located and sometimes immersed in "$\Lambda$" background field or cloud.

(6) We may be able to obtain information from other spherical universe, especially its information in very early epoch.